\documentstyle[preprint,aps,tighten]{revtex}
\newcommand{\be}{\begin{equation}}
\newcommand{\ee}{\end{equation}}
\newcommand{\br}{\begin{eqnarray}}
\newcommand{\er}{\end{eqnarray}}

\newcommand{\half}{\frac{1}{2}}
\def\a{{\alpha}}
\def\b{{\beta}}
\def\g{{\gamma }}

\def\m{{\mu}}
\def\n{{\nu}}

\begin{document}
\draft
\title{Georgi-Goldstone Realization of Chiral Symmetry in Quantum Chromodynamics }
\author{R.Acharya ($^{\ast}$ )}
\address{Department of Physics and Astronomy, Arizona State University, Tempe AZ 85287}
\maketitle

\begin{abstract}

It is shown that quantum chromodynamics based on asymptotic freedom {\it and} confinement exhibits the {\it vector} mode of chiral symmetry conjectured by Georgi.

\end{abstract}

\vspace{4.1in}

\noindent ($\ast $): e-mail address: raghunath.acharya@asu.edu\\
\noindent \pacs{PACS numbers: 11.30 Qc., 11.30.Rd., 12.38.Aw.}

I shall begin by summarizing the key results constituting  conventional wisdom in standard Quantum Chromodynamics (QCD). 

Firstly, in vector-like gauge theories and in QCD in particular, non-chiral 
symmetries  such as $SU_{L+R}(2) \subset SU_L(2) \times SU_R(2)$  or $SU_{L+R} 
(3)
\subset  SU_L(3) \times SU_R(3)$  cannot be spontaneously broken. This 
is the Vafa-Witten result \cite{vafa,rapn1}.  In QCD with large $N_c$, 
Coleman and Witten \cite{CW} showed that chiral symmetry is broken to diagonal $U(N_F)$ 
and thus, if chiral symmetry is broken, it must happen in such a manner that 
flavor symmetry is preserved. Secondly, chiral $SU_L(3) 
\times SU_R(3)$ symmetry  in QCD with massless $u, d, s$ quarks must be spontaneously 
broken. However,  it is difficult to show \cite{QFT} that chiral $SU_L(2) \times SU_R(2) 
$ symmetry in QCD with massless $u, d$ quarks  is also spontaneously broken. 
Thirdly, QCD may very well  exhibit the Higgs mode for the vector current and 
the Goldstone mode for the axial vector current {\it i.e., \/} the massless 
scalars arising from Goldstone theorem get `eaten up' by the gauge vector 
field which consequently acquires a finite mass. This important conjecture was 
introduced by Georgi in 1989 who called it as a new realization of chiral 
symmetry (``vector mode" ) since it involves both the Wigner-Weyl and 
Nambu-Goldstone modes.  This is Georgi's conjecture \cite{Georgi}. 

To quote Weinberg \cite{Salam}, ``A recent paper of Georgi, can be 
interpreted as proposing that QCD at zero temperature is near a second order 
phase transition, at which the broken chiral $SU_L(3) \times SU_R(3)$ symmetry has 
a $(8,1) + (1,8)$ representation, consisting of the octet of pseudoscalar 
Goldstone bosons plus an octet of massless scalars, that on the broken symmetry 
side of the phase transition, become the helicity-zero states of the massive 
vector meson octet,  $\cdots$.   It is intriguing and mysterious that at the second 
order phase transition at which  chiral $SU_L(2) \times SU_R(2)$ of massless 
QCD becomes {\it unbroken \/}, this symmetry may become {\it local\/} with $\rho$ and $A_1$ as 
massless gauge bosons".

As a 
final point one may observe that the charges corresponding to spontaneously 
broken local gauge symmetries are {\it screened\/} and the vector mesons are 
massive. This is a manifestation of spontaneously broken local symmetries. For 
instance, the well-known example \cite{Higgs} is the Abelian Higgs model: in 
the spontaneously broken phase, the vector field has a finite mass (thus the 
field is of finite range) and the conserved current does not have a total 
charge in the physical Hilbert space. 

The nature of spontaneously broken chiral symmetry \cite{tHooft} is intimately connected to spontaneously broken scale invariance. It has been emphasized by Adler \cite{AdlerRMP} that there are two examples of relativistic field theories which exhibit spontaneously broken scale invariance where chiral symmetry is also broken. These are Johnson-Baker-Wiley model of quantum electrodynamics  
\cite{BJW} and asymptotically free gauge theories. This indeed may be a 
general feature as we pointed out recently in our investigation \cite{rapn2}
of spontaneously broken chiral symmetry in QCD. Let us review this connection 
briefly.

Unbroken scale invariance can be expressed as
\be
Q_D(t) |0> \;  = 0,
\ee
where the dilatation charge is 
\be
Q_D(t)= \int \; d^3x \; D_0( {\bf x}, t),
\label{2}
\ee
defined in terms of $D_{\m}( {\bf x},t )$, the dilatation current. 
Equivalently,
\be
\partial^{\m} D_{\m} |0>=0.
\ee
Invoking Coleman's theorem    \cite{Coleman}, which is valid for 
continuous symmetries,  we can then prove that the divergence of the 
dilatation current itself must  vanish identically:
\be
\partial^{\m} D_{\m}=0.
\ee 
On the other hand, it turns out that we know that the divergence of the dilatation current is 
determined by the trace anomaly \cite{Collinsetal}  in QCD:
\be
\partial^{\m}D_{\m}= \half \; \frac{\b(g)}{g} G^{\a}_{\m \n} G^{\m \n}_{\a} + 
\sum_i \; m_i [1 + \g_i (\theta)] \bar{\psi}_i  \psi_i,
\label{5}
\ee
where the second term vanishes for massless quarks in the chiral limit. 
Consequently the beta function must vanish. It is well-known that in an asymptotically free theory 
of QCD which also exhibits confinement, the behavior of $\b(g)$ is such that 
it decreases as $g$ increases and never turns over. Consequently $g=0$ is the 
only possibility and hence the theory reduces to triviality. We therefore 
conclude by {\it reductio ad absurdum\/} that scale invariance must be broken 
spontaneously by the QCD vacuum state
\be
Q_D(t)|0>  \; \neq 0 \, .
\ee
Thus, scale invariance is broken {\it both} ``spontaneously" by the vacuum state and explicitly by the trace anomaly. Consequently, the states obtained by successive repeated application of $Q_D(t)$ on the vacuum state are neither vacuum states nor are they necessarily degenerate.

Let us now consider the commutator \cite{rapn2}
\be
[Q_D (0),Q_a (0) ]= -i d_Q Q_a (0)
\label{7}
\ee
which defines the scale dimension of the charge $Q^a(0)$, the generators of 
vector $SU(N)$ (the flavor (non-chiral) group of QCD). Eq.(\ref{7}) may be ``promoted" to arbitrary time by introducing the operator $e^{iHt}$ on the left, $e^{-iHt}$ on the right (and  inserting the unit operator  $1= e^{-iHt} \;
e^{iHt}$ in the middle of the commutator on the left hand side):
\be
[Q_D(t),  Q_a(t)] = - i d_Q Q_a(t)\, .
\label{8}
\ee
It is 
important to point out that operator relations such as Eq.(\ref{8}) are 
{\it unaffected\/} by spontaneous symmetry breaking, which is manifested in 
the properties of physical {\it states\/}, as emphasized by Weinberg 
\cite{QFTagain}.

I shall now proceed to establish that the flavor vector charges, $Q_a(t)= \int \; d^3x \; V_a^0( {\bf x}, t) $, are screened, {\it i.e.,} $Q_a(t)=0$, where $V_a^{\m}$ are the conserved vector currents in QCD. ($a= N_F^2 -1$, where $N_F$ is the number of flavors). Since $V_a^{\m}$ are conserved, the corresponding vector charges $Q_a(t)$ commute with $H$:
\be
[Q_a(t), \, H ]=0,
\label{9}
\ee
where $H$ is the Hamiltonian (density), $H= \Theta^{00}$. Eq.(\ref{9}) implies that the following double commutator also vanishes:
\be
[Q_D(t), \, [Q_a(t),\, H]\, ]=0,
\label{10}
\ee
where $Q_D(t)$ is the dilation charge defined in Eq.(\ref{2}). Let us now invoke the Jacobi identity to recast Eq.(\ref{10}):
\be
[Q_a(t), \, [H,\, Q_D(t)\, ]\, ] + [H, [ Q_D(t), Q_a(t)] \,] =0.
\label{11}
\ee
Since
\be
[H, \, Q_D(t)] = -i \partial_{\m }D^{\m} ( {\bf x},t) \not=0,
\label{12}
\ee
by virtue of the trace anomaly, Eq.(\ref{5}), and the second double commutator on the left hand side of Eq.(\ref{11}) vanishes in view of Eqs.(\ref{8},\ref{9}), we arrive at the important {\it operator} relation:
\be
[Q_a(t), \, \partial^{\m}D_{\m} ( {\bf x}, t ) \, ] =0 \, .
\label{13}
\ee
Applying Eq.(\ref{13}) on  the vacuum state, we obtain:
\be
[Q_a(t), \, \partial^{\m}D_{\m} ( {\bf x}, t ) \, ] |0 \rangle =0 \, .
\label{14}
\ee
We now invoke the Vafa-Witten result \cite{vafa} 
\be
Q_a(t) |0 \rangle=0.
\label{15}
\ee
From Eqs.(\ref{14},\ref{15}), we conclude \cite{note1} that
\be
{\cal O}( {\bf x}, t) | 0 \rangle \equiv Q_a(t) \partial^{\m}D_{\m} ( {\bf x }, t ) \,  |0 \rangle =0 \,
\label{16}
\ee
where the operator $\cal O$ is {\it local} in space and time \cite{note2}. Consequently we can utilize the all-powerful Federbush-Johnson theorem which applies to any local operator to conclude that 
\be
Q_a(t) \partial^{\m}D_{\m} ( {\bf x }, t ) \,  \equiv 0 \,
\label{17}
\ee
where $\partial^{\m}D_{\m} ( {\bf x }, t )$ is governed by the trace anomaly,   \cite{note2}.

Since $\partial^{\m}D_{\m} ( {\bf x}, t )$ {\it cannot} vanish in QCD exhibiting both asymptotic freedom {\it and} confinement (except at $g=0$ ), we are led to conclude that the vector charges are screened, {\it i.e.,} $Q_a(t) =0$, proving Georgi's \cite{Georgi} conjecture: QCD at zero temperature exhibits chiral symmetry in the Nambu-Goldstone mode ($N_F = 3$) and the vector symmetry is realized in the Higgs mode (``vector mode") in the sense conjectured by Georgi, {\i.e.,} the vector charges $Q_a$ are screened and the corresponding vector mesons become massive by devouring the would-be scalar Goldstone bosons which disappear from the physical spectrum. The key ingredients in this analysis are the Vafa-Witten theorem \cite{vafa}, the Federbush-Johnson theorem \cite{Coleman} and the trace anomaly for the divergence of the dilation current \cite{Collinsetal}. Some concluding remarks are in order.

First of all, Eq.(\ref{9}) which is the {\it local} version expressing current conservation $\partial^{\m}V_{\m}^a =0$ holds if the surface terms at infinity can be neglected. This is, {\it a posteriori}, justified since the vector charges $Q_a(t)$ must annihilate the vacuum (Vafa-Witten theorem) and hence the flavor vector symmetry {\it cannot} be spontaneously broken ({\it i.e.,} there are no scalar Goldstone bosonsto produce a long range interaction which would have resulted in a non-vanishing surface term).  Secondly, it may be worthwhile amplifying the connection between the vanishing of the vector charges, {\it i.e.,} $Q_a(t)=0 $ and Georgi's conjecture. Let us consider a massive vector boson coupled to a conserved source (by construction). While in the massless case, this is mandatory, in the massive case, one can choose the current to be conserved if one so wishes. In such a case, clearly the charges can be non-vanishing, even though the corresponding particle is massive. But the essential point here is that if the charges vanish, $Q_a(t)=0$, the vector meson  cannot be massless: it must be massive ! This is the desired connection with Georgi's conjecture.

Finally, there are unresolved issues with the vector limit advocated by Georgi, such as its compatibility with lattice results and the question as to why aren't the pions also eaten up by the axial vector $A_1$'s ? These issues remain to be resolved, perhaps, in a future publication.

I am indebted to P. Narayana Swamy for numerous  conversations on the perennial topic of chiral symmetry and for our previous efforts in trying to nail down Georgi's conjecture.


\begin{thebibliography}{30}
\bibitem{note1}
It is worth pointing out that it would be fallacious to conclude from Eq.(\ref {16}) that $\partial^{\m}D_{\m} \, |0\rangle =0$.
\end{thebibliography}
\end{document}